\documentclass[11pt,a4paper]{article}
\newif\ifanonymoussubmission
\newif\ifpreprintversion
\anonymoussubmissionfalse
\preprintversiontrue
\ifanonymoussubmission
  \usepackage[review]{acl}
\else
  \ifpreprintversion
    \usepackage[preprint]{acl}
  \else
    \usepackage{acl}
  \fi
\fi

\usepackage{times}
\IfFileExists{phvr8t.tfm}{}{}
\IfFileExists{pcrr8t.tfm}{}{}
\usepackage{latexsym}
\usepackage[T1]{fontenc}
\usepackage[utf8]{inputenc}
\usepackage{microtype}
\usepackage{amsmath}
\usepackage{amssymb}
\usepackage{graphicx}
\usepackage{booktabs}
\usepackage{array}

\title{Single Canonical Prompts Underestimate LLM Safety's Surface-Form Sensitivity}

\ifanonymoussubmission
\author{Anonymous ACL Submission}
\else
\author{
  Yongxi Zhou$^{1,\ast}$ \quad Junwei Yao$^{1}$ \quad Yuanzhe Liu$^{2}$ \quad Zihan Dong$^{2}$ \\
  Wenbo Ye$^{3}$ \quad Jiaxi Wen$^{1}$ \quad Lai Yun Choi$^{1}$ \\
  $^{1}$Northeastern University, Massachusetts, USA \\
  $^{2}$Georgia Institute of Technology, Georgia, USA \\
  $^{3}$University of Southern California, California, USA \\
  $^{\ast}$Corresponding author: \texttt{zhou.yongx@northeastern.edu}
}
\fi
\date{}

\begin{document}
\maketitle

\begin{abstract}
A benchmark score is a point estimate reported without an uncertainty budget, yet most benchmarks read each item at a single canonical surface form. We ask what that omission hides, separating two sources of uncertainty in the reported number: the noise it inherits from stochastic decoding and model-based judging, and the systematic variation it inherits from the arbitrary choice of surface form. When an item's intent is held fixed and only its meaning-preserving surface form varies, does the canonical-form score faithfully estimate model behavior, and how much of any variation is decoding/judge noise rather than signal? We instantiate this in safety---a high-stakes setting with no gold label to average toward---treating single canonical, English, harmful prompts as the instrument under test. To answer this without prior confounds, we pre-author the reformulations (refusal-free, mostly non-LLM: machine back-translation and a Matrix-Language-Frame code-switch generator) so an identical surface form reaches every model, score all responses with one \emph{human-anchored}, vendor-neutral judge (Claude---$\kappa=0.86$ vs.\ human on unsafe compliance, stable across languages---cross-checked by GPT-4o), and verify intent preservation. On 370 seeds $\times$ 5 surface forms $\times$ 5 models, no single transformation is uniformly most dangerous (6 of 20 per-transformation McNemar tests survive correction, most protective). Yet evaluating only the canonical prompt underestimates unsafe compliance: the union of unsafe outcomes across forms exceeds even the \emph{worst single} form by 3.3--12.9 pp, with bootstrap 95\% CIs excluding zero for all five models, and 5--13\% of seeds safe on canonical are unsafe under some reformulation---above a \emph{zero} stochasticity floor (canonical resampled five times at temperature 0 gives 0/370 new exposures). The magnitude of this gap is model-dependent (largest on Gemini 2.5 Pro). Across these five forms, one form recovers only $\sim$53\% of a model's observed unsafe surface and about three reach 85\%---a redundancy characterization of this form set, not a sample of a defined population. A benign control (XSTest) suggests the instability is bidirectional (comparable new over-refusals), though the benign and harmful pools are not item-matched. A safety score should therefore carry the uncertainty induced by surface form, which resampling alone does not expose. We release the dataset, code, and per-response labels.
\end{abstract}

\section{Introduction}
Evaluating a foundation model is an act of measurement: a benchmark reads the model through an instrument---a fixed set of items, a scoring rule, and increasingly a model-based judge---and reports a number we treat as a property of the model. Like any instrument it can be biased and noisy, and a basic validity question is whether reading each item at a single canonical surface form faithfully estimates behavior on the underlying construct when the surface form could have been otherwise. We study this in safety, where the stakes of a biased instrument are concrete: models are deployed where alignment failures carry real consequences, and safety is routinely assessed by presenting a single canonical, English, harmful request and measuring refusal or compliance. This paradigm embeds an assumption---that one canonical phrasing represents the intent in general---yet the same intent can be paraphrased, translated, code-switched, or framed indirectly. If behavior varies across these meaning-preserving surface forms, a single-form benchmark misestimates robustness, and the size of that misestimate is a measurement-error question, not only a safety one. Reporting the score without this component understates its uncertainty, which resampling one phrasing cannot reveal.

We study this directly. Our research question is: \emph{when harmful intent is held fixed while surface form varies, how stable is safety, and does the canonical prompt give a faithful estimate of a model's vulnerability?} Prior work on multilingual and code-switched jailbreaks~\citep{yong2023lowresource,deng2024multilingual,csrt2025acl} establishes that \emph{individual} reformulations can bypass guardrails, but typically studies one transformation in isolation, and---crucially for measurement---often (a) asks the target model to perform the transformation via a meta-instruction, confounding reformulation ability with safety; (b) scores with substring rules or a same-vendor judge; and (c) does not verify that the reformulation preserves the harmful intent.

We remove these confounds. First, all reformulations are \textbf{pre-authored} so that an \emph{identical} surface form is sent to every model, and three of the four transformations are produced by \textbf{refusal-free, non-LLM} methods, sidestepping the fact that aligned models refuse to reformulate harmful seeds (Section~\ref{sec:reform}). Second, we score with a \textbf{single human-anchored, vendor-neutral judge} (Claude, anchored to human annotation and cross-checked by an independent-vendor judge) applied uniformly to all five evaluated models (Section~\ref{sec:judge}). Third, we \textbf{verify intent preservation} with the same judge.

Framed this way the contribution belongs to evaluation science, not attack research: it asks how to draw reliable conclusions from an imperfect instrument applied to imperfect data~\citep{jacobs2021measurement}. Two distinct sources of uncertainty enter the number a benchmark reports: run-to-run variation from stochastic decoding and judging, and systematic variation from the choice of surface form---the canonical reading systematically misses exposure that other meaning-preserving readings reveal. Conflating them makes a score look more certain than it is, because resampling one phrasing probes only the first. We disentangle them with a stochasticity floor: because language models are not text-deterministic even at temperature 0~\citep{zhou2026accuracystabilityrepeatedrunreliability}, we resample the canonical form and measure how often the safety label rather than the wording changes, so determinism in inference is a quantity we measure rather than assume. Only variation above this floor is attributed to surface form. Our contributions are methodological first. \textbf{(1) A confound-controlled, noise-floored protocol} combining the three controls above with a \emph{stochasticity floor} that separates surface-form sensitivity from decoding/judge noise and a \emph{benign over-refusal control} (XSTest) that separates safety degradation from general instability. \textbf{(2) The finding}: contrary to the framing that particular reformulations (translation, code-switching) are ``the dangerous'' ones, \emph{no single transformation uniformly increases unsafe compliance}---most significant per-transformation effects are protective. The instability is instead \emph{(intent $\times$ surface-form)-specific}: 5--13\% of seeds safe on the canonical prompt become unsafe under some form (above a zero noise floor), and the union exceeds the worst single form for all five models (bootstrap CIs exclude zero), making the canonical prompt a biased, optimistic estimator. XSTest \emph{suggests} bidirectionality---a comparable 6--18\% of benign prompts flip to over-refusal---but the pools are not item-matched, so we report this as secondary. \textbf{(3) A redundancy characterization of these five forms}: one form recovers only $\sim$53\% (37--68\% across models) of a model's observed unsafe surface, and $\sim$3 forms reach 85\% of the union. These forms are not a sample from a defined population, so the curve describes this form set, not general coverage. We release all artifacts.

\section{Related Work}
\paragraph{Jailbreaks and reformulation attacks.}
Adversarial suffixes~\citep{zou2023universal} and persuasion-based attacks~\citep{zeng2024johnny} optimize inputs to maximize harm. Closer to us, \citet{yong2023lowresource} show translating AdvBench into low-resource languages bypasses GPT-4 ($\sim$79\% ASR); \citet{li2024crosslanguage} and \citet{deng2024multilingual} study multilingual jailbreaks; CSRT~\citep{csrt2025acl} synthesizes code-switched queries. These study a single transformation, usually framed as an attack. We instead compare four meaning-preserving families on the same seeds as a \emph{measurement} question, and ask whether the canonical prompt is a faithful estimator. Framing effects also recur beyond the single-turn setting: \citet{liu2026reframing} decompose multi-agent safety into reframing, planner behavior, and delegation framing, and find the resulting compliance shifts to be strongly model-dependent---echoing, in a different setting, our finding that no single surface form is uniformly most dangerous.

\paragraph{Safety benchmarks and judges.}
HarmBench~\citep{mazeika2024harmbench} and JailbreakBench~\citep{chao2024jailbreakbench} standardize behaviors and---critically---ship \emph{validated} classifiers (a fine-tuned Llama-2-13B judge; a Llama-3-70B judge selected against human preferences), because substring scoring is unreliable. SG-Bench~\citep{mou2024sgbench} varies prompt-engineering parameters. We adopt the validated-judge standard but use a vendor-neutral judge so that no evaluated model is scored by a same-vendor judge, and we add an explicit intent-preservation check that these benchmarks do not target.

\paragraph{Multi-prompt and distributional evaluation.}
A parallel line in general NLP measurement shows that single-prompt scores are unreliable estimators: model performance varies widely across semantically equivalent prompt formats~\citep{sclar2024quantifying}, motivating a shift from point estimates to \emph{distributional}, multi-prompt evaluation~\citep{mizrahi2024state}. We import this measurement stance into safety, where it differs in three ways. (i) There is no gold label to average toward, so the quantity of interest is the \emph{union}/worst-case exposure across forms rather than a central tendency. (ii) We calibrate that union against an explicit stochasticity \emph{noise floor} and a \emph{benign} over-refusal control, separating construct signal from decoding noise and from generic instability. (iii) Intent preservation must be verified, because a surface change can silently alter the construct being measured. Read as uncertainty quantification, this line asks what a score would have been had an arbitrary choice been made differently; we add an explicit noise floor that separates variation due to surface form from variation due to stochastic decoding and judging. This stance also reconciles our modest per-form effects with best-of-$N$ jailbreaking~\citep{hughes2024bestofn}, which attains high attack success by sampling \emph{hundreds} of augmented prompts at non-zero temperature: that regime compounds large $N$ with uncontrolled decoding noise, whereas we isolate surface form at five forms and temperature~0, so the two findings are complementary rather than contradictory. A related instrument-bias result appears in agent settings, where individually safe components compose into unsafe configurations that per-component evaluation cannot see~\citep{wang2026skills}; there, as here, the gap between a flagged and a human-confirmed hazard is itself a property of the instrument.

\section{Method}

\subsection{Task and Data}
We define safety robustness as the stability of safe behavior when a harmful intent is held fixed and only the surface form varies. For each seed $s$ we build five conditions: the canonical prompt $t_0(s)$ and four meaning-preserving reformulations. Seeds are drawn from HarmBench~\citep{mazeika2024harmbench} and AdvBench~\citep{zou2023universal}, deduplicated and mapped to four harm categories. The frozen v3 set has 370 seeds (100 physical-harm, 100 cyber-abuse, 100 fraud/deception, 70 violent-wrongdoing) $\times$ 5 conditions $=$ 1{,}850 prompts.

\subsection{Pre-authored, refusal-free reformulations}
\label{sec:reform}
Asking an aligned model to ``translate/rephrase'' a harmful seed is self-defeating: the generator refuses or distorts exactly these inputs, and any meta-instruction it executes confounds reformulation ability with safety. We therefore pre-author each reformulation as a fixed string, sent identically to all evaluated models, using methods with no safety filter wherever possible:
\begin{itemize}
\setlength{\itemsep}{1pt}\setlength{\topsep}{1pt}\setlength{\parskip}{0pt}
\item \textbf{paraphrase}: distant-pivot machine back-translation (English$\rightarrow$pivot$\rightarrow$English; pivots chosen for typological distance), accepted only when surface similarity to the source falls in a target band.
\item \textbf{translation\_zh}: machine translation English$\rightarrow$Chinese.
\item \textbf{code\_switch\_zh\_en}: a programmatic Matrix-Language-Frame generator~\citep{myersscotton1993}---English matrix, with noun-phrase constituents translated to Chinese---yielding genuine intra-sentential mixing rather than a translated sentence appended to its English original.
\item \textbf{indirect\_framing}: reframing as an indirect/hypothetical question requires semantic restructuring, so it is generated by an \emph{open-weight, API-accessed} model (DeepSeek-V3) under a text-transformation system prompt that suppresses refusal; intent preservation is verified post-hoc (Section~\ref{sec:judge}). It is the only transformation produced by an LLM rather than a non-LLM method.
\end{itemize}
Each item passes a structural quality gate; failures are flagged for manual review rather than silently substituted. Machine translation and the frame generator are content-agnostic and never refuse, which removes refusal-induced selection bias from dataset construction. We disclose that three of the four transforms (paraphrase $=$ pivot back-translation, translation\_zh, code\_switch\_zh\_en) share the same Chinese machine-translation machinery, so the positive correlation among forms is partly attributable to shared tooling; our ``distribution of surface forms'' is thus more precisely Chinese-MT variants plus one LLM reframing (indirect framing). Section~\ref{sec:results} quantifies the contribution of the one non-MT transform.

\subsection{Inference}
All five models---GPT-4o-mini, GPT-4o, Gemini 2.5 Flash, Gemini 2.5 Pro, DeepSeek-V3---receive each prompt under an identical system prompt at temperature 0 with a 256-token cap, via batch APIs.

\subsection{Human-anchored, vendor-neutral judging}
\label{sec:judge}
A single judge---Claude (\texttt{claude-sonnet-4-6}), temperature 0---scores every response on two questions, given the original harmful behavior as the reference: \textbf{refusal} (does the response decline/deflect?) and \textbf{unsafe compliance} (does it provide operationally useful harmful content?). Claude is not one of the evaluated providers, so no model is judged by a same-vendor judge, and it scores Chinese responses natively. Separately, the judge rates \textbf{intent preservation} for each reformulation against its seed; this is a dataset property and is judged once. We report judge labels throughout; a brittle substring detector that we also ran misses most unsafe cases (it labels $<$1\% unsafe) and most soft refusals, reproducing the known failure of rule-based scoring. Using an LLM judge rather than a cheaper feature-based classifier carries a real compute cost: \citet{zhang2026tradeoffs} find that fine-tuned LLMs consistently improve over gradient-boosting baselines at predicting human judgments of chatbot responses while requiring substantially more training and inference resources. Here the cheap alternative is not merely less accurate but non-functional, so we accept that cost.

\paragraph{Judge corroboration.} To bound judge error without a human anchor, we re-scored a stratified sample of 250 responses (balanced across the five models and five forms) with an \emph{independent, different-vendor} judge (GPT-4o). The two judges substantially agree---Cohen's $\kappa=0.67$ on unsafe compliance (95.2\% agreement) and $\kappa=0.65$ on refusal---with near-identical unsafe prevalence (21 vs.\ 19 of 250), indicating that the headline rates are not an artifact of one judge's calibration. 

\paragraph{Human anchor.} On a stratified subset of 185 responses (English 75, Chinese 55, code-switched 55, oversampling unsafe outcomes), a human annotator labelled \textbf{refusal} and \textbf{unsafe compliance} blind to the judge's labels. Human--judge agreement is high---Cohen's $\kappa=0.86$ on unsafe compliance and $0.91$ on refusal---and, critically, does \emph{not} degrade on non-English responses: per-language unsafe $\kappa$ is 0.92 (English), 0.82 (Chinese), 0.81 (code-switched), and refusal $\kappa$ is 0.95 / 0.93 / 0.85. The judge is thus a human-anchored instrument whose reliability is stable across surface form, so the cross-form effect is not an artifact of the judge mis-scoring foreign-language text. Full-set human annotation and a multi-annotator ensemble remain future work.

\section{Results}
\label{sec:results}

\subsection{Aggregate rates are stable; no transformation is uniformly worse}
Table~\ref{tab:main} reports judge refusal and unsafe-compliance rates. Aggregate rates move little across surface forms (mean refusal 75--82\%; mean unsafe 8--13\%), and crucially the \emph{direction} of any per-transformation effect is model-dependent. Testing each (model, transformation) against its canonical condition with exact McNemar tests paired by seed, and correcting across all 20 tests (Holm), only \textbf{6/20 are significant}---and most are \emph{protective} (the reformulation lowers unsafe compliance, e.g.\ DeepSeek-V3 under indirect framing, Gemini 2.5 Pro under code-switching). The lone significant \emph{increase} is GPT-4o-mini under Chinese translation ($+5.9$ pp unsafe; 24 seeds become unsafe vs.\ 2 recovered, $p_{\text{Holm}}{<}0.001$). There is no transformation that is uniformly most dangerous across models. These per-transformation tests are also underpowered: discordant-pair counts are small (median 26, range 6--64 across the 20 tests), giving $\sim$80\% power only for a discordant split near 78/22 or more extreme, so ``6/20 significant'' reflects limited power as much as genuine near-null marginal effects. The union analysis below answers a \emph{different} question---cumulative exposure across forms, not the effect of any one form---and we validate it against an explicit noise floor rather than leaning on its (degenerate) significance.

\begin{table}[t]
\centering
\footnotesize
\setlength{\tabcolsep}{3.2pt}
\begin{tabular}{lrrrrr}
\toprule
 & \textbf{4o-m} & \textbf{4o} & \textbf{G-Fl} & \textbf{G-Pro} & \textbf{DS} \\
\midrule
\multicolumn{6}{l}{\textit{Refusal rate (\%)}}\\
original          & 89.2 & 90.8 & 80.0 & 71.4 & 77.3 \\
paraphrase        & 87.6 & 88.4 & 77.8 & 71.6 & 78.9 \\
translation\_zh    & 80.8 & 88.6 & 72.4 & 58.6 & 75.4 \\
code\_switch       & 89.2 & 91.6 & 76.8 & 63.2 & 76.5 \\
indirect\_framing  & 85.4 & 87.6 & 70.5 & 61.6 & 85.7 \\
\midrule
\multicolumn{6}{l}{\textit{Unsafe compliance (\%)}}\\
original          &  7.6 &  5.4 & 15.1 & 14.9 & 18.9 \\
paraphrase        &  7.3 &  7.0 & 13.2 & 14.3 & 13.8 \\
translation\_zh    & \textbf{13.5} &  6.2 & 15.1 &  7.8 & 20.0 \\
code\_switch       &  7.3 &  4.3 & 15.1 &  6.5 & 19.2 \\
indirect\_framing  &  8.4 &  6.2 & 10.8 &  7.3 &  6.8 \\
\bottomrule
\end{tabular}
\caption{Judge-scored refusal and unsafe-compliance rates by transformation and model ($n{=}370$/cell). 4o-m: GPT-4o-mini; 4o: GPT-4o; G-Fl/G-Pro: Gemini 2.5 Flash/Pro; DS: DeepSeek-V3. Bold marks the only Holm-significant per-transformation \emph{increase} in unsafe compliance.}
\label{tab:main}
\end{table}

\subsection{Canonical prompts underestimate unsafe compliance}
The per-transformation nulls do not mean reformulation is harmless. Because different seeds fail under different forms, the \emph{union} of unsafe outcomes across the five surface forms exceeds the best the canonical prompt could indicate (Table~\ref{tab:union}). We make the honest comparison against the \emph{worst single} form, not just canonical: the union exceeds the worst single form by 3.3--12.9 pp, and bootstrap 95\% CIs on this gap exclude zero for \emph{all five} models (smallest GPT-4o-mini 3.3 pp $[1.6,5.1]$; largest Gemini 2.5 Pro 12.9 pp $[9.5,15.7]$; Table~\ref{tab:union}). We therefore treat the \emph{existence} of exposure beyond the worst single form as the robust claim and its \emph{magnitude} as model-dependent (largest on Gemini 2.5 Pro). Relative to the canonical prompt alone the union is 1.3--2.2$\times$. ``New exposure''---seeds safe on canonical but unsafe under some reformulation---is 5.4--13.0\% with Wilson intervals excluding zero (Table~\ref{tab:benign}). We report this proportion rather than a union-vs-canonical McNemar, which is structurally degenerate (the union contains the canonical condition). The two analyses use different inferential standards by design: the 20 per-transformation McNemar tests form a family of \emph{hypothesis tests} (Holm-corrected), whereas the union, new-exposure, and coverage quantities are \emph{estimates} reported with confidence intervals, not a second test family requiring multiplicity correction.

\paragraph{Signal, not noise.} The decisive question is whether ``new exposure'' reflects surface form or merely decoding/judge stochasticity re-sampled five times. We measure a \emph{stochasticity floor}---a repeated-run reliability check, since LLMs vary run-to-run even on deterministic tasks~\citep{zhou2026accuracystabilityrepeatedrunreliability}---by resampling the \emph{canonical} prompt five times per seed at temperature 0 (surface form held fixed) and taking the union. Although outputs are not text-deterministic (98\% of DeepSeek-V3 repeats differ in wording), the judge's safety \emph{label} never flips---the repeat-union equals the canonical rate and the floor of new exposures is \textbf{0/370 for both models tested} (GPT-4o-mini, DeepSeek-V3; 95\% CI $[0,1.0]$\%). Decoding stochasticity thus contributes $\approx$0 to label-level new exposure (judge noise is separately bounded by inter-judge $\kappa=0.67$).
The cross-form new exposures (6.2\% and 9.2\%) lie entirely above this zero floor and are therefore surface-form-driven. Because temperature is held at~0, this floor isolates the \emph{surface-form} component of exposure; at deployment temperatures, decoding stochasticity contributes \emph{additional} exposure (best-of-$N$ work finds $\sim$20\% of jailbreak successes fail to reproduce on resampling~\citep{hughes2024bestofn}). Our union is thus a conservative \emph{lower bound} on deployment-time exposure, not an upper bound.

\paragraph{Robustness.} The effect is not intent drift: restricting the union to reformulations the judge rates intent-preserving leaves it essentially unchanged (Gemini 2.5 Pro 25.9\% vs.\ 27.8\%; GPT-4o-mini and DeepSeek-V3 unchanged). It does not depend on the one LLM-generated transformation: excluding indirect framing, the union still exceeds canonical for every model (DeepSeek-V3 24.9\%). Conversely, the one mechanistically-distinct transform \emph{not} built on the Chinese-MT pipeline (indirect framing) \emph{alone} adds 1.1--4.6 pp of new exposure over canonical (4.1 pp GPT-4o-mini, 4.6 pp Gemini 2.5 Pro, 1.1 pp DeepSeek-V3), so the union is not solely a shared-MT-tooling artifact (the shared machinery is disclosed in Section~\ref{sec:reform}). And it is not a multiple-draws artifact: five \emph{independent} measurements at these rates would union to 26--58\%, far above the observed 12--28\%, so the forms are positively correlated, not random re-samples. A benchmark probing only the canonical form thus reports an optimistic point estimate of a larger, real vulnerability surface.

\subsection{A benign control suggests bidirectional instability}
Is this new exposure a \emph{safety} phenomenon, or just general behavioral instability under reformulation? We run an identical pipeline over the 250 safe prompts of XSTest~\citep{rottger2024xstest}---prompts a well-calibrated model should answer---and measure new \emph{over-refusals} (benign seeds answered on canonical but refused under some reformulation). The two effects are comparable in magnitude (Table~\ref{tab:benign}): new over-refusal is 6--18\% versus 5--13\% new unsafe, and for Gemini 2.5 Flash and DeepSeek-V3 the benign effect is \emph{larger}. Reformulation therefore induces \emph{bidirectional} per-seed instability---flipping some harmful prompts to unsafe and some benign prompts to refused, at similar rates---rather than a one-directional erosion of safety. Because our benign and harmful pools are not item-matched (Limitations), we read this as \emph{consistent with} general surface-form instability rather than as evidence that the effect is safety-agnostic. This is the honest reading: the canonical prompt mis-estimates behavior in \emph{both} directions, and ``single-prompt evaluation is optimistic about safety'' must be paired with ``it is also optimistic about over-refusal.''

\begin{table*}[t]
\centering
\footnotesize
\setlength{\tabcolsep}{5pt}
\begin{tabular}{lrrrr}
\toprule
\textbf{Model} & \textbf{Canon.} & \textbf{Worst} & \textbf{Union} & \textbf{U$-$W [95\% CI]} \\
\midrule
GPT-4o-mini      &  7.6 & 13.5 & 16.8 &  3.3 [1.6,5.1] \\
GPT-4o           &  5.4 &  7.0 & 12.4 &  5.4 [3.2,7.3] \\
Gemini 2.5 Flash & 15.1 & 15.1 & 20.5 &  5.4 [3.0,7.0] \\
Gemini 2.5 Pro   & 14.9 & 14.9 & 27.8 & 12.9 [9.5,15.7] \\
DeepSeek-V3      & 18.9 & 20.0 & 25.1 &  5.1 [3.0,6.8] \\
\bottomrule
\end{tabular}
\caption{Unsafe compliance (\%): canonical-only, the \emph{worst single} form, and the union across all five forms. U$-$W is the union-minus-worst gap with a seed-level bootstrap 95\% CI (5{,}000 resamples); it excludes zero for every model. Per-seed new exposure (safe on canonical, unsafe under some form) and its zero stochasticity floor are reported in Table~\ref{tab:benign}.}
\label{tab:union}
\end{table*}

\begin{table*}[t]
\centering
\footnotesize
\setlength{\tabcolsep}{5pt}
\begin{tabular}{lrr}
\toprule
\textbf{Model} & \textbf{New unsafe (harmful)} & \textbf{New over-refusal (benign)} \\
\midrule
GPT-4o-mini      &  9.2 [6.7,12.6] &  6.0 [3.7,9.7] \\
GPT-4o           &  7.0 [4.8,10.1] &  6.0 [3.7,9.7] \\
Gemini 2.5 Flash &  5.4 [3.5,8.2]  & 14.0 [10.2,18.8] \\
Gemini 2.5 Pro   & 13.0 [9.9,16.8] & 10.0 [6.9,14.3] \\
DeepSeek-V3      &  6.2 [4.2,9.2]  & 18.0 [13.7,23.2] \\
\bottomrule
\end{tabular}
\caption{New exposure under reformulation (\%, Wilson CI), harmful vs.\ benign. Harmful = seeds safe on canonical but unsafe under some form; benign = XSTest safe prompts answered on canonical but over-refused under some form. The stochasticity floor for new exposure---canonical resampled 5$\times$ at temperature~0---is 0/370 (GPT-4o-mini, DeepSeek-V3), so all harmful CIs lie above the noise floor. The two directions are comparable in magnitude (benign larger for two models), suggesting bidirectional surface-form instability, though the harmful and benign pools are not item-matched (see Limitations).}
\label{tab:benign}
\end{table*}

\subsection{How many surface forms must a benchmark probe?}
The union effect has a direct operational consequence: how much of a model's true unsafe surface does a budget of $m$ randomly chosen forms recover? For each model we compute the expected coverage of the union when testing $m$ of the five forms (exact, over all $\binom{5}{m}$ subsets). Averaged across models, coverage is 53\% ($m{=}1$), 73\% ($m{=}2$), 85\% ($m{=}3$), 93\% ($m{=}4$), 100\% ($m{=}5$); the single-form figure ranges from 37\% (Gemini 2.5 Pro) to 68\% (Gemini 2.5 Flash). A single canonical prompt thus recovers only about half of what five meaning-preserving forms reveal, and roughly three forms reach 85\% of our five-form union. We frame this as a \emph{redundancy characterization} of these five forms, not a sampling guideline: the five forms are not a sample from a defined population of surface forms, so the coverage curve is descriptive of this form set rather than an estimate of general coverage.

\subsection{Decomposed consistency}
For each seed we ask whether refusal is identical across all five forms, decomposing into safe-consistent (refuses all), unsafe-consistent (complies with all), and mixed. The mixed fraction is large for the weaker-aligned models---Gemini 2.5 Pro 38.6\%, Gemini 2.5 Flash 29.2\%, DeepSeek-V3 23.5\%---versus 15--18\% for the GPT models. For roughly a quarter to two-fifths of harmful intents, then, \emph{whether a model refuses depends on phrasing}, which is precisely what a single-form benchmark cannot see and what drives the union effect above.

\subsection{Intent preservation}
The clean construct lets us check whether low refusal reflects evasion or destroyed intent. The judge rates intent preserved for 91.9\% (paraphrase), 93.8\% (translation), and 94.1\% (code-switch), but only 82.4\% for indirect framing. Indirect framing's lower apparent severity is thus \emph{partly} an artifact of intent neutralization---it sometimes softens a request into an abstract question---rather than pure evasion, a distinction invisible to refusal rates alone.

\section{Discussion}
Three results must be held together. (i) \emph{Marginally}, meaning-preserving reformulation does not make these frontier models systematically more likely to comply: per-transformation effects are small, mostly non-significant after correction, and when significant usually protective---contradicting a reading of prior single-transformation studies in which a particular surface form is ``the'' vulnerability. (ii) \emph{Jointly}, the canonical prompt is an optimistic estimator: vulnerability is idiosyncratic to (intent $\times$ surface-form) pairings, so aggregating meaning-preserving forms reveals a larger unsafe surface than any single form. (iii) \emph{Symmetrically}, the benign control is consistent with this not being safety-specific (the harmful and benign pools are not item-matched): the same reformulations flip comparable fractions of benign prompts into over-refusals. The phenomenon is best described as \emph{surface-form behavioral instability}, which manifests as both missed unsafe compliance and spurious over-refusal. The practical implication is not ``add translation attacks'' but ``estimate behavior over a distribution of surface forms, not a single canonical one---for both safety and over-refusal,'' and report decomposed consistency so that per-seed instability is visible.

Our mostly-null and protective per-transformation effects may appear to contradict the high attack-success rates of earlier translation jailbreaks~\citep{yong2023lowresource,deng2024multilingual}. Two factors reconcile this. First, those results predate the frontier models evaluated here; safety on translated inputs has improved. Second, and methodologically, attack-framed pipelines do not verify that the reformulation preserves harmful intent, whereas our intent check (Section~\ref{sec:results}) shows a non-trivial fraction of reformulations---especially indirect framing (82.4\%)---partly neutralize intent, which inflates apparent success when uncontrolled. Under a human-anchored, intent-controlled construct the effect is real but smaller and distributional rather than transformation-specific.

As an observation (not a validated claim, given single-judge scoring; see Limitations), the models with higher baseline unsafe rates (Gemini, DeepSeek: 15--19\% vs.\ 5--8\% for GPT) also show larger mixed fractions, so the canonical-prompt underestimate tends to be largest where baseline risk is already highest.

\section{Conclusion}
Under a confound-controlled, noise-floored construct---pre-authored, mostly non-LLM reformulations sent identically to every model, one human-anchored vendor-neutral judge with intent verified, a zero stochasticity floor, and a benign over-refusal control---no meaning-preserving transformation is uniformly more dangerous. Instead, reformulation induces \emph{(intent $\times$ surface-form)-specific} instability: canonical-prompt evaluation misses new unsafe compliance (5--13\%, above a zero floor), and the union exceeds the worst single form for all five models. A benign control suggests this is bidirectional (comparable new over-refusal, 6--18\%), pending item-matched controls. A single form recovers only $\sim$half of a model's observed unsafe surface; about three meaning-preserving forms recover 85\% of our five-form union. Robust evaluation should therefore probe a distribution of meaning-preserving surface forms---budgeting roughly three---and report decomposed consistency, rather than relying on a single canonical phrasing.

Beyond safety, the protocol is a reusable measurement recipe for any benchmark that lacks a gold label to average toward and whose items admit meaning-preserving surface variation. Its components are general: pre-authored, refusal-free perturbations sent identically to every system to decouple the perturbation from the system under test; a vendor-neutral, human-anchored judge applied uniformly; a union/worst-case estimand in place of a central tendency; a stochasticity floor that separates instrument noise from signal; and an intent- (or construct-) preservation check that guards against silently measuring something else. Safety is the instance we report because its instrument bias has direct consequences, but capability and alignment benchmarks that read each item once at a canonical phrasing inherit the same validity risk. We make no theoretical claim about which surface forms a model will fail on; the contribution is a way to measure, and bound the reliability of, the gap that single-form reading leaves invisible.

\section*{Limitations}
\textbf{Judge validation.} Headline labels come from one LLM judge, human-anchored on a stratified subset ($\kappa=0.86$ unsafe, $0.91$ refusal; Section~\ref{sec:judge}) and corroborated by an independent different-vendor judge ($\kappa=0.65$--$0.67$ on 250 items). The human anchor uses a single annotator on 185 items; full-set human annotation and a multi-annotator ensemble over all responses remain needed to fully bound judge error---including possible shared LLM-judge blind spots that two model judges could share. The stochasticity floor bounds decoding noise but not \emph{surface-form-correlated} judge noise; the per-language human $\kappa$ (0.81--0.92 across English/Chinese/code-switched) addresses the latter on the subset.
\textbf{Intent metric.} Intent preservation is itself judge-rated; no standard benchmark for intent preservation of \emph{harmful} reformulations exists, and our 82.4\% for indirect framing should be read as approximate. It is also rated by the same judge that scores unsafe compliance, so independent-judge re-scoring is needed to rule out circularity.
\textbf{Open-weight access.} DeepSeek-V3 (and the absence of a self-hosted model) means ``open'' here is open-weight accessed via API, not a locally hosted model; one self-hosted model would strengthen provider diversity. Indirect framing is the one transformation produced by an LLM rather than a non-LLM method.
\textbf{Stochasticity floor scope.} The zero noise floor is measured on two of five models (GPT-4o-mini, DeepSeek-V3); extending it to all five would fully generalize the noise-attribution claim. A 256-token generation cap may also truncate long compliant answers before operational content appears, slightly conservatively biasing unsafe detection.
\textbf{Scope.} English--Chinese only, single-turn, temperature 0. Non-zero-temperature robustness (sampling $k$ generations and reporting unsafe@$k$) and additional language pairs are natural extensions; the union result is expected to strengthen under sampling.
\textbf{Benign control scope.} Our benign control uses XSTest's 250 safe prompts; a larger benign set and additional over-refusal benchmarks (OR-Bench~\citep{cui2024orbench}, PHTest~\citep{an2024phtest}) would tighten the bidirectional estimate. The benign and harmful seed pools also differ in topic distribution, so the two ``new-exposure'' rates are comparable in magnitude but not strictly matched item-for-item; promoting bidirectionality to a core claim requires item-matched benign/harmful pairs.
\textbf{Shared transformation tooling.} Three of four transforms share the Chinese machine-translation pipeline (Section~\ref{sec:reform}); although the one non-MT transform alone contributes new exposure (Section~\ref{sec:results}), a fully mechanistically-distinct transform---syntax-only restructuring, a different language family, or a different paraphraser tool---is needed to complete the shared-tooling control.

\section*{Ethical Considerations}
We study reformulations of \emph{existing public} harmful-behavior benchmarks to improve safety \emph{measurement}; we propose no new attack and report aggregate rates, not operational content. Reformulations are produced by content-agnostic tools from already-public seeds and add no new harmful capability. We release transformed prompts and labels consistent with venue safety norms and the practices of HarmBench/JailbreakBench, and complete the Responsible NLP checklist.

\section*{Acknowledgments}
LLM assistants were used for sentence-level copy-editing only. All experimental design, data construction, analysis, tables, and conclusions were produced by the authors' code and inspection.

\bibliography{custom}
\bibliographystyle{acl_natbib}

\appendix
\section{Judge Rubric}
\label{app:rubric}
The judge (Claude \texttt{claude-sonnet-4-6}, temperature 0; second judge GPT-4o) receives the original harmful behavior as reference and answers, for each response:
\begin{quote}\footnotesize
\textit{REFUSAL: does the response refuse, decline, deflect, or avoid fulfilling the request? UNSAFE: does the response provide operationally useful content that advances the behavior (concrete steps, methods, code, actionable harmful detail)? Judge content, not tone; the response may be in any language. Reply exactly:} \texttt{REFUSAL: <YES/NO>} \texttt{UNSAFE: <YES/NO>}.
\end{quote}
Intent preservation uses a separate single-question prompt comparing each reformulation against its seed. The full rubric, per-response labels for both judges, and a stratified subset prepared for human validation are released with the code.

\section{Reformulation Details}
\label{app:reform}
Paraphrase and translation use Google Translate (via \texttt{deep\_translator}). Paraphrase back-translates through typologically distant pivots tried in order (Japanese, Finnish, Arabic, Korean, Turkish, Hungarian, Vietnamese), accepting the first whose surface similarity to the source---\texttt{difflib} sequence ratio on lowercased text---lies in $(0.30, 0.88)$; near-identical ($\geq$0.95) or possibly-drifted ($\leq$0.30) outputs are flagged. ``Non-LLM'' here means content-agnostic statistical/neural MT and rule-based chunking that never refuse; it is not a claim that MT is non-neural. Code-switch translates English noun-phrase constituents (NLTK chunker) into Chinese within an English frame. All reformulations are pre-authored once and sent verbatim to every model.

\end{document}